\documentstyle[preprint,aps]{revtex}
\title{Dynamics of the swelling or collapse
of a homopolymer}
\author{E.Pitard \thanks{e-mail:pitard@spht.saclay.cea.fr} and H.Orland \\
Service de Physique Th\'eorique \\
Centre d'Etudes de Saclay, CEA, 91191 Gif-sur-Yvette  \\
France} 

\date{\today}
\begin{document}
\maketitle

\begin{abstract}

We study the dynamics of a polymer
when it  is quenched from a $\theta$ solvent into a good
or bad solvent by means of a Langevin equation.
The variation of the radius of gyration is studied
as a function of time.
For the first stage of collapse 
or swelling, the characteristic time-scale is found to be independent of 
the number of
monomers.
Other scaling laws are derived
for the diffusion regime at larger times.
Although the present model is solved 
only for homopolymers and doesn't
include hydrodynamic interactions, these results
may be a first step towards the understanding
of the early stages of protein folding.

\end{abstract}

\def\be{\begin{equation}}
\def\ee{\end{equation}}
\def\bea{\begin{eqnarray}}
\def\eea{\end{eqnarray}}
\def\bea*{\begin{eqnarray*}}
\def\eea*{\end{eqnarray*}}
\def\bit{\begin{itemize}}
\def\eit{\end{itemize}}

Under certain conditions (such as a temperature decrease or
exposure to a bad solvent), 
homopolymers undergo a  collapse (coil-globule) transition.
This has been studied quite extensively,
and it has regained some interest since it may be the 
simplest  model to describe the first stages of protein folding.
Few experimental data are available \cite{CHU}\cite{BILL}, 
but a lot of theoretical
work and computer simulations have been done, leading to a large diversity of 
results.
Although the 
thermodynamic properties of the homopolymer problem have been quite well understood,
there is at present no analytical derivation of the kinetics laws 
for the polymer collapse.
Since one might expect the first stages of protein folding to
be dominated by non-specific hydrophobic forces, this
phenomenon should be quite similar to the collapse of a homopolymer
chain.
It is thus highly desirable to
develop a general analytical technique in order to study the kinetics
of polymer collapse.

According to de Gennes' theory \cite{PGG}, the process of collapse of a
flexible coil leads to the formation of crumples on a minimal scale
along the linear chain, which thickens and shortens
under diffusion of the monomers, then
forms new crumples of growing scale,
until the final state of a compact globule
is reached; the total time of collapse is
estimated as 
$$\tau_c = {\eta a^3 \over k_B \theta } {|\Delta T| \over \theta} N^2$$

where $\eta$ is the viscosity of the solvent, $\theta$ is the $\theta$
temperature, $a$ is the monomer size and $\Delta T$ is the temperature
quench. This time $\tau_c$ has a very strong dependence on molecular
weight. For the case of proteins (see reference \cite{BILL}), $N=300$,
which yields a collapse time of $\tau_c \sim 1 \ \mu s$.

In a revised model, A.Buguin et al. \cite{BUG} consider a different 
mechanism, in two steps: first a fast formation
of "pearls" along the chain, followed by a slower stage 
of compaction.

Grosberg et al. \cite{GRO} have included the role of self entanglement
to de Gennes' model, and consider a two stage mechanism:
first a collapse leading to a crumpled globule
with characteristic time $\sim N^{2}$,
and then a chain knotting driven by reptation
with a longer characteristic time
$\sim N^{3}$.

Monte Carlo studies (Ostrowsky$\&$Bar-Yam \cite{OST},
 Milchev$\&$Binder \cite{MIL})
and Langevin dynamics simulations (Byrne et al. \cite{BYR})
lead to diverging interpretations concerning the
mechanisms of collapse.

In a series of articles, Timoshenko  et al.
\cite{TIM1}
have developed an alternative theory based on 
a self-consistent method using Langevin equations
that can be analyzed numerically; kinetics laws for the collapse
of a homopolymer are obtained with or without 
hydrodynamics, at early and later stages.

Experimentation in this field is quite difficult
since one has to work in a very dilute regime
in order to avoid aggregation of chains
and observe a collapse.
Experiments by Chu et al. \cite{CHU} on polystyrene in cyclohexane
reveal a two-stage kinetics of approximately equal times.
The most promising experiments are those by
Chan et al. \cite{BILL}  who study sub-millisecond protein folding
by ultra-rapid mixing;
based on optical techniques, these experiments 
can monitor folding
up to the microsecond time-scale.

In the following, we shall present an analytical method to study
the kinetics of a homopolymer in a $\theta$ solvent when it is quenched
 into good solvent conditions (swelling into a coil) or into bad solvent
conditions (collapse into a globule).

We consider a homopolymer chain in a $\theta$ solvent - i.e a Gaussian coil-
consisting of N monomers, obeying  the Langevin dynamics as 
the chain is quenched 
into good or bad solvent conditions (equations (\ref{1}) and (\ref{2}) ).

Neglecting hydrodynamical interactions (the Oseen tensor) the
equations of motion for the system read:

\begin{eqnarray}
    && \frac{\partial{r}}{\partial{t}} = -\Gamma_{0} 
        \frac{\partial{H}}{\partial{r}} + \eta(s,t) \label{1}\\
    && H=\frac{1}{a_{0}^{2}} \int_{0}^{N}
        \left(\frac{\partial{r}}{\partial{s}}\right)^{2}ds +V(r(s,t)) \label{2}
\end{eqnarray}

where $N$ is the total number of monomers, 
$r(s,t)$ is the position of monomer $s$ in the chain,
$a_{0}$ is  the monomer length and $\Gamma_{0}={D \over k_B T}$
where $D$ 
is the diffusion constant
of a monomer in the solvent and $k_B T$ is the temperature. The
intra-molecular as well as intermolecular interactions of the
chain are contained in the potential $V(r(s,t)) $.
The thermal noise $\eta(s,t)$ is a Gaussian noise with zero
mean and correlation given by:
$$<\eta(s,t)\eta(s,t')>=2D\delta(t-t')$$

The method consists in finding a virtual homopolymer chain
which obeys a simpler Langevin equation,
chosen so that its
radius of gyration best approaches the radius of
gyration of the real chain at each time $t$.

The virtual chain, defined by $r^{(v)}(s,t)$ satisfies the Langevin equation:

\begin{eqnarray}
    && \frac{\partial{r^{(v)}}}{\partial{t}} = -\Gamma_{0} 
        \frac{\partial{H_{v}}}{\partial{r^{(v)}}} + \eta(s,t) \label{3}\\
    && H_{v}=\frac{1}{a^{2}(t)} \int_{0}^{N}
       \left (\frac{\partial{r^{(v)}}}{\partial{s}}\right)^{2}ds  \label{4}
\end{eqnarray}

with the same diffusion constant and noise as the original equation,
but with a much simplified Hamiltonian $H_v$. Indeed this Hamiltonian
$H_v$ represents  a Gaussian chain, but with a time dependent  Kuhn
length $a(t)$.

Our method is a generalization of Edwards' 
Uniform Expansion Model \cite{EDW} to dynamics. This method consists
in 
calculating the radius of gyration
of a polymer in a good solvent by using perturbation theory, and
adjusting the simplified Hamiltonian so that the first order
perturbation to the radius of gyration vanishes.
If $v$ denotes the excluded volume,
the method gives the Flory radius \cite{Flory} for large $N$
and agrees with the result of the first-order perturbation 
expansion for small $v$. Note that it would seem natural to use
the most general quadratic Hamiltonian rather than that of (\ref{4})
, but this was shown by des Cloizeaux
\cite{desCloizeaux} to yield the incorrect exponent $\nu = 2/d$.

Let's define

$$ \chi(s,t)=r(s,t)-r^{(v)}(s,t)$$
$$W= H - H_v$$

Assuming that (\ref{3}) is a good approximation
to (\ref{1}) , $\chi(s,t)$ and $W$ can be regarded as small,
and to first order in these quantities, the dynamical equations
become:

\begin{eqnarray}
    && \frac{\partial{r^{(v)}}}{\partial{t}} = 
        \frac{\Gamma_{0}}{a^{2}(t)} 
            \frac{\partial^{2}{r^{(v)}}}{\partial{s^{2}}} +\eta(s,t) \label{5}\\
   && \frac{\partial{\chi}}{\partial{t}} = 
         \frac{\Gamma_{0}}{a^{2}(t)} 
            \frac{\partial^{2}{\chi}}{\partial{s^{2}}}
+\Gamma_{0}\left[\left(\frac{1}{a_{0}^{2}}-
      \frac{1}{a^{2}(t)}\right)\frac{\partial^{2}{r^{(v)}}}{\partial{s^{2}}}
  +F(r^{(v)}(s,t))\right] \label{6}
\end{eqnarray}

where $F(r(s,t))=-\frac{\partial{V}}{\partial{r(s,t)}}$
is the driving force for the swelling or collapse of the chain.

In this approximation, the radius of gyration of the chain
becomes:

\begin{eqnarray}
  R_{g}&&=\frac{1}{N}\int_{0}^{N}<r^{2}(s,t)>ds  \\
    &&\simeq \frac{1}{N}\int_{0}^{N}<((r^{(v)})^{2}(s,t) + 2 r^{(v)}(s,t) \chi(s,t))>ds  
\label{7}
\end{eqnarray}

The brackets denote the thermal average ( that is an average over the
Gaussian noise $\eta(s,t)$).
Our approximation consists in choosing the parameter $a(t)$ in such a
way that the first order in (\ref{7}) vanishes:

\begin{equation}
   \int_{0}^{N}<r^{(v)}(s,t)\chi(s,t)>=0 \label{8}
\end{equation}

or in Fourier coordinates:

\begin{equation}
    \sum_{n\neq 0}<\tilde{r}^{(v)}_{n}(t)
                    \tilde{\chi}_{n}^{*}(t)>=0  \label{9}
\end{equation}

where the Fourier transform is given by:

 \begin{eqnarray*}  
&& \left\{\begin{array}{ll}  
                 \tilde{r}_{n}(t)=\frac{1}{N}
         \int_{0}^{N}e^{i\omega_{n}s}r(s,t)ds\\  
                 r(s,t)=\sum_{n\neq 0}
      e^{-i\omega_{n}s} \tilde{r}_{n}(t) 
          \end{array} \right. \\ 
 \end{eqnarray*}   
We have used periodic boundary conditions, so that 
$\omega_{n}=\frac{2\pi n}{N}$.
In addition, to get rid of the center of mass diffusion, we constrain
the center of mass of the system to remain at fixed position,
$\tilde{r}_{0}(t)=
 \tilde{r}^{(v)}_{0}(t)
=\tilde{\chi}_{0}(t)=0$.

Equations (\ref{5}) and (\ref{6}) can easily be solved in Fourier
space.
We assume that at time $t=0$, the chains are in a $\theta$ solvent, so
that the initial condition $\{r(s,0)\}$ obeys Gaussian statistics. We
choose the initial virtual chain to coincide with the real one, so
that $r^{(v)}(s,0) = r(s,0)$ for any $s$. Denoting by
$\overline{\cdots}$ the average over the initial conditions,
the correlation function of $r(s,0)$ (in Fourier space) is taken as:

\begin{eqnarray}
\overline{\tilde{r}_{n}(0)} &&= 0 \\
\overline{\tilde{r}_{m}(0)\tilde{r}_{n}^*(0)}
  && =\frac{Na_{0}^{2}}{4\pi^{2}n^{2}}\delta_{mn} 
\end{eqnarray}

Replacing 
$\tilde{r}^{(v)}_{n}(t)$ and
$ \tilde{\chi}_{n}^*(t)$ by their expression
in (\ref{9}), and taking thermal and initial condition 
averages, we obtain an implicit equation for $a(t)$.


This equation can be solved analytically in both
limits
$ t << \tau_R $ (short time limit) and $ t >> \tau_R $
(long time limit)
where $\tau_R = {N^2 a_0^2 \over 4 \pi^2 D}$
is the Rouse time.

In the following,
we neglect all hydrodynamic interactions
and consider only two-body and three-body
interactions between monomers.

More precisely, for a chain in a good solvent,
we consider excluded volume interactions:

 $$
 V(r(s,t))=V_{2}(r(s,t))=
       \frac{v}{2}\int_{0}^{N}ds\int_{0}^{N}ds'
        \delta(r(s,t)-r(s',t))    
 $$

and for a chain in a bad solvent,
we take attractive two-body interactions and repulsive
three-body interactions:

 \begin{eqnarray*}
&& V(r(s,t))=-V_{2}(r(s,t)) +V_{3}(r(s,t))\\
&& V(r(s,t))=-\frac{v}{2}\int_{0}^{N}ds\int_{0}^{N}ds'
        \delta(r(s,t)-r(s',t))\\
 &&  +\frac{w}{6}\int_{0}^{N}ds\int_{0}^{N}ds'\int_{0}^{N}ds''
        \delta(r(s,t)-r(s',t)) \delta(r(s',t)-r(s'',t)) , 
 \end{eqnarray*} 

where $v>0$ and $w>0$.

In good solvent conditions, at short times $t << \tau_R$, the radius
of gyration increases like a power law, which we recast in the form of
a stretched exponential:

\begin{equation}
R_{g}^{2}(t)=Na_{0}^{2}e^{\sqrt{\frac{t}{\tau_{c}}}},
\end{equation}
where the characteristic time $\tau_c$ is defined below in (\ref{tauc}).

For large time $t>> \tau_R$, the radius of gyration relaxes to its
Flory value:

\begin{equation}
R_{g}(t)\sim N^{\frac{3}{d+2}}a_{0}(1-e^{-\frac{t}{\tau_{1}}}),
\end{equation}

with a relaxation time $\tau_1$ given by:
$$\tau_{1}\sim N^{\frac{d+8}{d+2}} $$

Note that this relaxation time is much larger than the Rouse time, for
dimensions lower than 4. For example in $d=3$,  $\tau_{1}\sim
N^{\frac{11}{5}}$.

In a bad solvent, at short times $t << \tau_R$, the radius
of gyration decreases as above:

\begin{equation}
R_{g}^{2}(t)=Na_{0}^{2}e^{-\sqrt{\frac{t}{\tau_{c}}}},
\end{equation}

with a characteristic time $\tau_c$ defined in (\ref{tauc})
and for large times $t>> \tau_R$, the radius of gyration relaxes to
that of compact globule according to

\begin{equation}
R_{g}(t)\sim (\frac{w}{v})^{\frac{1}{d}}
            N^{\frac{1}{d}}(1-e^{-\frac{t}{\tau_{2}}}),
\end{equation}

where

$$\tau_{2}\sim (\frac{w}{v})^{\frac{2}{d}} N^{1+\frac{2}{d}}. 
$$ 

Note that for dimensions larger than $2$, 
the relaxation time is much shorter than the Rouse time.
For example in $d=3$,  $\tau_{2}\sim
N^{\frac{5}{3}}$.

In the two cases discussed above, the dynamic relaxation exponent
$z$, relating the relaxation time of the system to its radius of gyration:
\be
\tau \sim R_g^z
\ee
satisfies the exponent relation:
\be 
z=2+{1 \over \nu}
\ee
This relation was derived by de Gennes \cite{PGG2}.

The first stage of the swelling or collapse can be
described by a stretched exponential with
characteristic time $\tau_{c}$
given by:

\begin{equation}
\tau_{c}^{\frac{1}{2}}=
\frac{2}{3}\frac{\pi^{\frac{d}{2}}a_{0}^{d+1}}
     {\sqrt{2D \pi}vI_{d}N^{\frac{4-d}{2}}}
\label{tauc}
\end{equation}

where
$$
I_{d}=\int_{0}^{1}du
        \int_{0}^{1}du'
       \frac{1}{[|u-u'|(1-|u-u'|)]^{\frac{d}{2}}}  \  \  \ {\rm with}
\ \ |u-u'| > \Lambda$$

and $\Lambda=1/N$ is a short distance cut-off.

For $d<2$, the integral converges for small 
$\Lambda$, and $I_d$ is independent of $N$.
On the other hand, 
for $d\geq2$, the integral is infra-red divergent and 
thus there is an
explicit dependence on the cut-off.
It is easily seen that this $N$ dependence exactly cancels
out the $N$ dependence in (\ref{tauc}) so that the final
characteristic time $\tau_c$ is finite (independent of $N$).
In particular, for $d=3$, 
we find
$$\tau_{c}\sim \frac{2\pi^{2}}{9 D}
    \left(\frac{a_{0}^{3}}{v}\right)^{2}
      \left({a_{0} \over 5.22}\right)^{2}$$.

The typical order of magnitude of this short time collapse can be
calculated for a typical protein in water. The diffusion constant
of an amino-acid in water is typically
$D\sim 10^{-7}cm^{2}/s$
. Taking a monomer length of $a_{0}\sim 4 \AA$, for a chain of 100 aminoacids,
we find a microscopic characteristic time
$\tau_{c}\sim 10^{-9}s$ (in these conditions, the Rouse time
is $\tau_R \sim 4. 10^{-6}$ s). Note that in these conditions, the relaxation
time $\tau_2$ is of the order of magnitude of the Rouse time.
This time is several orders of magnitude lower
than other estimates in the literature ( see references
\cite{thirum}, \cite{BILL} ). 
The fact that this characteristic time is independent of $N$ shows
that in the early stage, the collapse is a very local phenomenon,
where nearby monomers aggregate into small domains.

Whereas previous mechanisms relied
mainly on pure diffusion,
we have introduced here 
explicit attractive hydrophobic forces for the collapse 
of the chain.
These forces are determinant at the early stages of the process, and
make it much faster than pure diffusion.

At larger time scales, we find other scaling laws for the relaxation
towards equilibrium (which is either
the collapsed globule or the swollen coil
within the Flory theory).
Let us mention that 
we find the same relaxation time for the collapse
as in Kuznetsov et al. \cite{TIM1}, namely
( $\tau_{2}\sim N^{\frac{5}{3}}$). This time is several orders of
magnitude larger than $\tau_c$

Although in this work we did not take into account the role of
hydrodynamic interactions, we expect their effect
to be weak, at least  at the early
stage of the collapse.
However, these effects might 
become important in later stages of folding.

\acknowledgements{
We thank S.Doniach and B.Eaton for suggesting us 
the study of the early stages of protein folding.

We thank T.Garel for a critical reading of the manuscript.
}
\bibliographystyle{unsrt}

\end{document}